\documentclass[aps,prl,reprint, superscriptaddress]{revtex4-2}
\usepackage{bbold}
\usepackage{comment}
\usepackage[colorlinks, citecolor=red]{hyperref}
\usepackage[utf8]{inputenc}
\usepackage[T1]{fontenc}    %
\usepackage[english]{babel}
\usepackage[pdftex]{graphicx}
\usepackage{amsmath,amssymb,amsthm,bbm, mathtools} %
\usepackage{mathrsfs}
\usepackage{xcolor}
\usepackage[normalem]{ulem}
\usepackage{multirow}
\usepackage{braket}

\begin{document}

\title{Fast readout of quantum dot spin qubits via Andreev spins}
\author{Michèle Jakob}
\altaffiliation{These authors contributed equally to this work.}
\affiliation{QuTech and Kavli Institute of Nanoscience, Delft University of Technology, Delft, The Netherlands}
\author{Katharina Laubscher}
\altaffiliation{These authors contributed equally to this work.}
\affiliation{Condensed Matter Theory Center and Joint Quantum Institute, Department of Physics, University of Maryland, College Park, Maryland 20742, USA}
\author{Patrick Del Vecchio}
\affiliation{QuTech and Kavli Institute of Nanoscience, Delft University of Technology, Delft, The Netherlands}
\author{Anasua Chatterjee}
\affiliation{QuTech and Kavli Institute of Nanoscience, Delft University of Technology, Delft, The Netherlands}
\author{Valla Fatemi}
\affiliation{School of Applied and Engineering Physics, Cornell University, Ithaca, NY, 14853, USA
}
\author{Stefano Bosco}
\email{s.bosco@tudelft.nl}
\affiliation{QuTech and Kavli Institute of Nanoscience, Delft University of Technology, Delft, The Netherlands}

\begin{abstract}
Spin qubits in semiconducting quantum dots are currently limited by slow readout processes, which are orders of magnitude slower than gate operations. In contrast, Andreev spin qubits benefit from fast measurement schemes enabled by the large resonator couplings of superconducting qubits but suffer from reduced coherence during qubit operations. Here, we propose fast and high-fidelity measurement protocols based on an electrically-tunable coupling between quantum dot and Andreev spin qubits. In realistic devices, this coupling can be made sufficiently strong to enable high-fidelity readout well below microseconds, potentially enabling mid-circuit measurements. Crucially, the electrical tunability of our coupler permits to switch it off during idle periods, minimizing crosstalk and measurement back-action. Our approach is fully compatible with germanium-based devices and paves the way for scalable quantum computing architectures by leveraging the advantages of heterogeneous qubit implementations.
\end{abstract}

\maketitle

\def\thefootnote{*}\footnotetext{These authors contributed equally to this work.}

\paragraph{Introduction.--}
Quantum dot spin qubits (DSQs)~\cite{PhysRevA.57.120}, especially those based on silicon and germanium (Ge) heterostructures~\cite{Scappucci2021,RevModPhys.95.025003}, are leading candidates for large-scale quantum computing~\cite{Stano2022}. These systems combine a compact footprint and compatibility with industrial semiconductor fabrication processes with high quantum performance~\cite{Neyens2024,maurand2016cmos,Piot2022,camenzind2021spin,bosco2023phase,PRXQuantum.2.010348,Petit2020,Xue2021,Zwerver2022,George2025,steinacker2024300,Jirovec2021,Yoneda2018,Takeda2022,Liles2024,Borsoi2023}. DSQs exhibit long coherence times and support high-fidelity single- and two-qubit gate operations, often exceeding fault-tolerance thresholds~\cite{Xue2022,doi:10.1126/sciadv.abn5130,Noiri2022,Lawrie2023}, with gate times of tens of nanoseconds~\cite{hendrickx2020fast,geyer2022two,doi:10.1126/science.ado5915,carballido2024qubit,bassi2024optimal,Froning2021,Wang2022} and operation demonstrated in mid-scale devices~\cite{john2024two,Philips2022,hendrickx2020four,Zhang2025}.
Despite these advantages, a key challenge in DSQ systems remains the readout, which is significantly slower than gate operations because of small DSQ-readout channel couplings~\cite{10.1063/5.0088229}. State-of-the-art devices require integration times  typically of tens to hundreds of microseconds, to reach high-fidelity spin readout~\cite{Stano2022,PhysRevX.8.021046,PhysRevX.13.011023,PhysRevApplied.13.024019,Zheng2019,Urdampilleta2019,Zhao2019}. This readout bottleneck  introduces idling errors and poses serious limitations for quantum error correction protocols and algorithms requiring frequent measurements.

\begin{figure}
\centering
\includegraphics[scale=0.42]{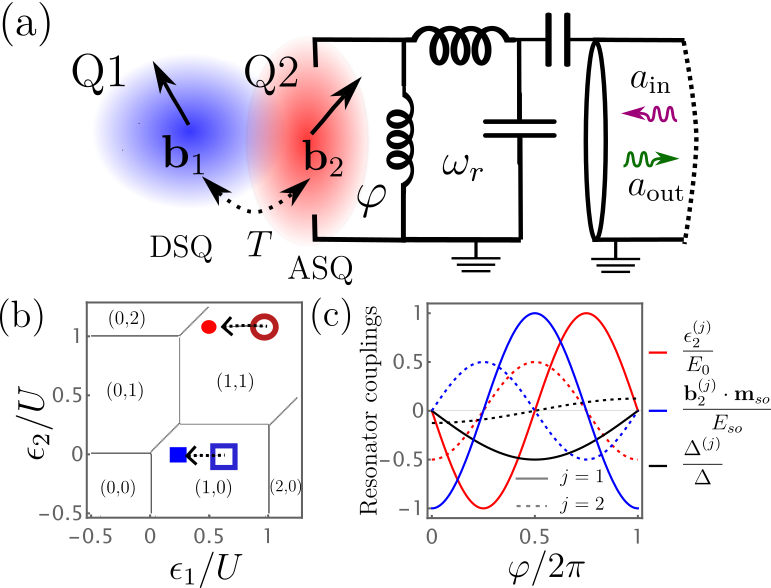}
\caption{\textbf{Tuneable ASQ-DSQ coupler.} (a)  A DSQ (Q1) is tunnel-coupled to an ASQ (Q2), inductively coupled to a microwave resonator.  (b) Charge stability diagrams for $U_{11}=U_{22}=4U_{12}\equiv U$ at $T=\Delta=0$, see Eqs.~\eqref{eq:HDQD} and~\eqref{eq:HT}. Filled blue and red dots represents two fast-readout points with large ASQ-DSQ hybridization when Q2 is empty and occupied, respectively. The hybridisation is suppressed electrically by decreasing $T$ or modifying the Q1-Q2 detuning by moving towards the hollow dots, where high-fidelity quantum gates on DSQs can be performed without back-action from the ASQ. (c) Amplitude of the inductive resonator-ASQ interactions against  $\varphi$, see Eq.~\eqref{eq:Hint}. 
\label{fig:1}
}
\end{figure}

In contrast, Andreev spin qubits (ASQs) comprising spins localized in semiconductors proximitized by superconducting Josephson junctions~\cite{PhysRevLett.90.226806,PhysRevB.96.125416,  doi:10.1126/science.abf0345, Hays2020,Pita-Vidal2023,Pita-Vidal2024,PhysRevLett.131.097001,PRXQuantum.3.030311,PhysRevX.9.011010,lu2025andreev,hoffman2025decoherence} benefit from strong inductive coupling to on-chip microwave resonators, enabling high-fidelity readout within hundreds of nanoseconds~\cite{doi:10.1126/science.abf0345,lu2025andreev} and long-range connectivity~\cite{Pita-Vidal2024}.
Ge heterostructures are particularly promising for ASQs~\cite{Lakic2025,PRXQuantum.5.030357}. Their compatibility with superconductors~\cite{PhysRevB.108.155433,babkin2024superconducting,johannsen2025anomalous,pino2024theory,PhysRevB.107.035435,PhysRevB.109.035433} has recently enabled significant proximity-induced superconducting gaps~\cite{Lakic2025,Tosato2023,PhysRevResearch.3.L022005,Steele2024} and  demonstrations of gate-tuneable Josephson junctions~\cite{Sagi2024,Kiyooka2025}. The potential for isotopic purification~\cite{https://doi.org/10.1002/adma.202305703} and p-type wavefunction~\cite{PhysRevB.78.155329,PhysRevLett.127.190501,PhysRevB.101.115302} of Ge can suppress nuclear spin noise, likely a key limiting factor of current ASQs~\cite{Hays2020,Pita-Vidal2023}. Furthermore, the strong, tunable spin-orbit interaction (SOI) in hole nanostructures, especially large in lattice-matched unstrained~\cite{Desamaterial,PhysRevB.104.115425,mauro2025hole} and tensile-strained~\cite{del2025fully,PhysRevB.107.L161406,PhysRevB.110.045409} Ge channels, offers efficient electrical control of the qubit. These properties make Ge an ideal platform for hybrid superconducting-semiconducting spin qubit architectures~\cite{Leblanc2025,Valentini2024,PhysRevB.109.085303}.

Here, we propose a tunable coupling scheme between DSQs and ASQs, see Fig.~\ref{fig:1},  combining the high-fidelity gate operations of DSQs with the fast, efficient readout capability of ASQs.  The coupling is electrically switched: it remains off during DSQ gate operations to suppress crosstalk and back-action~\cite{goan2001continuous,PhysRevA.79.013819,svastits2025readout}, and is activated only during the readout phase. We present several readout protocols tailored to different experimental configurations, all compatible with Ge-based hybrid devices~\cite{Lakic2025,Sagi2024}. Our analysis indicates that readout fidelities beyond $99.9\%$ can be achieved well below microseconds, representing orders-of-magnitude improvements in measurement speed. We envision that by leveraging strong inductive ASQ-resonator coupling, which enables long-range ASQ-ASQ connectivity~\cite{PRXQuantum.6.010308,lu2024kramers}, our ASQ–DSQ coupler could also mediate entangling gates of distant DSQs, overcoming their short-range connectivity limitations~\cite{Vandersypen2017} and offering opportunities for flexible layouts in large-scale spin-based quantum processors.

\paragraph{Tunable ASQ-DSQ coupler.--}
We consider two quantum dots, where dot $i=1$ hosts a DSQ and dot $i=2$ is  proximitized by two superconducting leads with phase difference $\varphi$ and forms an ASQ, see Fig.~\ref{fig:1}(a). The coupled DSQ-ASQ  system is described by the  Hamiltonian
\begin{equation}
\label{eq:HDQD}
H_{0}=\sum_{i} {\textbf{c}_{i}^\dagger h_{i} \textbf{c}_{i}}  - {\Delta_\varphi({c}_{2\uparrow}^\dagger c_{2\downarrow}^\dagger +{c}_{2\downarrow} c_{2\uparrow})} + \sum_{ij,\sigma\sigma'}\frac{U_{ij}}{2} n_{i\sigma}n_{j\sigma'}  \ ,
\end{equation}
where  $\textbf{c}_i=(c_{i\uparrow},c_{i\downarrow})$ are  fermion operators with density $n_{i\sigma}=c_{i\sigma}^\dagger c_{i\sigma}$, see the Supplemental Material (SM) for more details~\cite{SM}. 
The  Hubbard intra-dot  and inter-dot interactions  $U_{ii}$  and $U_{12}=U_{21}$, respectively, satisfy $U_{12}^2<U_{11}U_{22}$~\cite{RevModPhys.75.1}, and  define the charge stability diagrams sketched in Fig.~\ref{fig:1}(b).
The single-particle Hamiltonian of dot $i$, $h_i=-(\epsilon_i+2U_{ii}) \sigma_0+\textbf{b}_i\cdot\pmb{\sigma}/2$, includes dot energies $\epsilon_i$, Zeeman energies $\textbf{b}_i$, and a shift $2U_{ii}$ that cancels the $n_{i\sigma}^2=n_{i\sigma}$ contributions of the Hubbard energy. 
For the DSQ the dot energy coincides with its chemical potential, $\epsilon_1=\mu_1$, and the Zeeman energy $\textbf{b}_1=\mu_B \textbf{B} g_1$ is determined by the magnetic field $\textbf{B}$ and the tuneable dot $g$-tensor~\cite{PhysRevB.104.115425,PhysRevLett.131.097002,Wang2024,rimbach2024spinless,PhysRevB.103.125201}. For the ASQ, $\epsilon_2=\mu_2+E_0\cos \varphi$ depends on $\varphi$ via the Josephson energy $E_0$. The Zeeman energy $\textbf{b}_2= \mu_B \textbf{B} g_2- E_{so}\sin(\varphi) \textbf{m}_{so}$ also includes the additional phase-dependent SOI-contribution $E_{so}$~\cite{PhysRevLett.90.226806,PRXQuantum.3.030311,PhysRevLett.131.097001} aligned to the SOI vector $\textbf{m}_{so}$. 
The ASQ also inherits from the superconducting leads the proximitized correlations $\Delta_\varphi=\Delta \cos(\varphi/2)$~\cite{SeoaneSouto2024,PhysRevB.107.115407,PhysRevB.68.035105,PhysRevB.110.134506}.

Crucially, DSQ and ASQ are coupled by the tunneling $T/h\in[1-10]$~GHz. Gate potentials efficiently control $T$ on-demand~\cite{doi:10.1126/science.ado5915}, turning it on only during  readout. The interaction
\begin{equation}
\label{eq:HT}
H_{T}=-{T}
\left[\textbf{c}_{1}^\dagger \tilde{S}_{so} \textbf{c}_{2} +\textbf{c}_{2}^\dagger \tilde{S}_{so}^\dagger \textbf{c}_{1}\right]\ 
\end{equation}
 includes the SOI-induced spin-flip unitary evolution  $\tilde{S}_{so}= e^{i\tilde{\theta}_{so} \tilde{\textbf{n}}_{so}\cdot \pmb{\sigma}/2}$, characterized by a rotation angle $\tilde{\theta}_{so}$ around the SOI axis $\tilde{\textbf{n}}_{so}$, generally not aligned to $\textbf{m}_{so}$. 
 
In unstrained Ge channels where SOI is maximized~\cite{Desamaterial,PhysRevB.104.115425,mauro2025hole}, we expect  $E_{so}/h\in [10-500]$~MHz, $E_{0}/h\in [0.3-1]$~GHz, and $\Delta/h\in [4-10]$~GHz, see the SM~\cite{SM}, similar to current ASQ~\cite{Hays2020,PhysRevLett.131.097001}. For in-plane magnetic fields up to 200~mT, within the critical field of Ge-compatible superconductors~\cite{Lakic2025,Tosato2023,PhysRevResearch.3.L022005},  $\textbf{b}_i/h\in [0.1-1]$~GHz~\cite{john2024two,doi:10.1126/science.ado5915}, with controllable direction and amplitude by dot and strain engineering~\cite{PhysRevB.104.115425,PhysRevLett.131.097002}.

\paragraph{Inductive coupling to resonator.--}
Our readout is performed by an inductively-coupled resonator with energy $\hbar\omega_r a^\dagger a$.
Inductive coupling  shifts the phase $\varphi\to \varphi + X \varphi_r$, where $X=(a^\dagger+a)/\sqrt{2}$. The effective zero-point-phase $\varphi_r= p \sqrt{2\pi Z_r/R_q}\in [0.02-1]$ depends on the resonator impedance $Z_r\in[0.05-1]$~k$\Omega$, the superconducting resistance quantum $R_q=h/4e^2$~\cite{RevModPhys.93.025005}, and is rescaled by the participation ratio $p\in [0.1-1]$ accounting for the partial fraction of the resonator phase reaching the ASQ. 
At $Z_r=300$~$\Omega$ and $p=0.3$, $\varphi_r=0.16$. We restrict to low-impedance resonators ($Z_r\ll 1$~k$\Omega$), in contrast to capacitively coupled systems~\cite{Dijkema2025,yu2022strong,de2023strong,Janik2025,Landig2018,mi2018coherent,PhysRevLett.129.066801,PhysRevB.96.235434,PhysRevX.12.021026} where high-impedance resonators are required to enhance spin-photon coupling at the cost of losses and complexity.

The coupling Hamiltonian
\begin{equation}
\label{eq:Hint}
H_I= \sum_{j>0} \!\varphi_r^j X^{j}\! \left(\!\textbf{c}_2^\dagger\frac{ \textbf{b}_2^{(j)}\cdot\pmb{\sigma}-2\epsilon_2^{(j)}}{2}\textbf{c}_2 - \Delta^{(j)}{c}_{2\uparrow}^{\dagger}  {c}_{2\downarrow}^{\dagger} +\text{h.c.}\! \right) 
\end{equation}
includes resonator-induced fluctuations of ASQ dot and Zeeman energy, $\epsilon_2^{(j)}= \partial^j_\varphi\epsilon_2/j\propto E_0$ and  $\textbf{b}_2^{(j)}= \partial^j_{\varphi} \textbf{b}_2/j\propto E_{so} \textbf{m}_{so}$, respectively, and the resonator-mediated pairing $\Delta^{(j)}=\partial^j_{\varphi} \Delta_\varphi/j\propto \Delta$.
The first and second order susceptibilities are shown in Fig.~\ref{fig:1}(c). The spin-independent terms $\epsilon_2^{(j)}$ and $\Delta^{(j)}$ are odd (even) functions of $\varphi$ at $j=1$ ($j=2$), while $\textbf{b}_2^{(j)}$ has the opposite parity. 

\paragraph{Single-particle regime.--}
We consider a single particle confined in our hybrid double dot, a stable configuration because of $U_{ij}$, see Fig.~\ref{fig:1}(b). To minimize cross-talk and ASQ-mediated noise, we assume DSQ quantum gates are performed in the (1,0) region (hollow blue square) and  ASQ-DSQ hybridisation is activated only during readout by increasing $T$ and moving closer to the (0,1) region (solid blue square).

From Eqs.~\eqref{eq:HDQD} and~\eqref{eq:HT} and using third-order perturbation theory, we find the hopping Hamiltonian
\begin{equation}
\label{eq:H_hopping_1P}
H_\text{1P}= -\tau_z\frac{\epsilon_\varphi}{2}+ \tau_1\frac{b_1\sigma_z}{2}+\tau_2\frac{b_2\sigma_z}{2} -T_\varphi ({\tau_+ {S}_{so} +\tau_- S^\dagger_{so} } ) \ ,
\end{equation}
where the Pauli matrices $\tau$ act on the dot subspace, $\tau_\pm=(\tau_x\pm i\tau_y)/2$ and $\tau_{1,2}=(\tau_0\pm \tau_z)/2$.
We aligned both Zeeman fields to the $z$-direction by a local spin-rotation $R_i$ generated by  $\tau_i U_i= \tau_i e^{-i \Theta_i \textbf{n}_i\cdot \pmb{\sigma}/2} $, that satisfies $U_i^\dagger (\textbf{b}_i\cdot\pmb{\sigma}) U_i= (R_i \textbf{b}_i)\cdot\pmb{\sigma} =b_i\sigma_z   $. In this frame, the spin-flip tunneling operator transforms as $ {S}_{so}= U_1^\dagger \tilde{S}_{so} U_2\equiv e^{i {\theta}_{so} \textbf{n}_{so}\cdot \pmb{\sigma}/2}$, with modified $\theta_{so}$ and $\textbf{n}_{so}$.

The correlations $\Delta_\varphi$ renormalize dot detuning $\epsilon_\varphi\approx \epsilon+\Delta_\varphi^2/ \mathcal{U}$ and tunneling $T_\varphi=T[1-\Delta_\varphi^2/(2\mathcal{U}(\mathcal{U}-\epsilon))]$, adding a $\varphi$ dependence to $T$. Here, $\epsilon=\epsilon_1-\epsilon_2$ and $\mathcal{U}=U_2+2U_{12}-2\epsilon_2$.
Finally, the phase-modulation of $\textbf{b}_2$ imprints a $\varphi$-dependence to $S_{so}$. We envision that this could enable efficient  ASQ-DSQ interactions engineering by biasing $\varphi$, also mediating additional resonator couplings that we neglect here.

\begin{figure}
\centering
\includegraphics[scale=0.45]{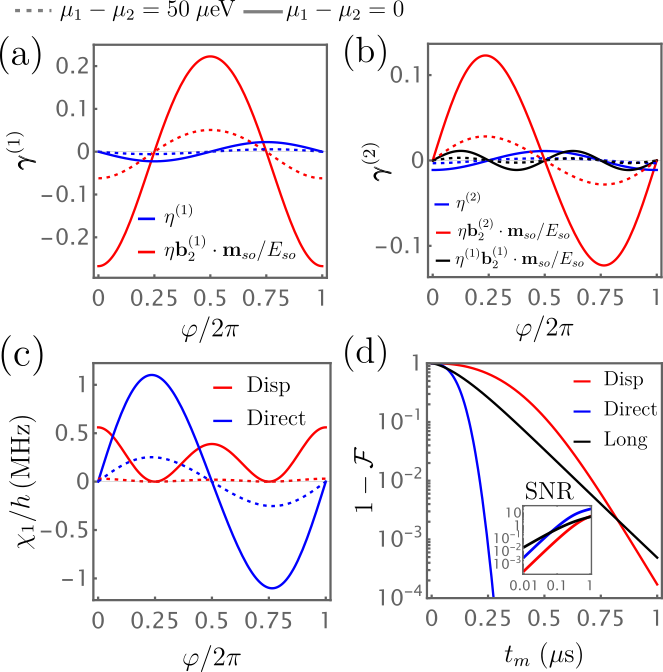}
\caption{ \textbf{Hybrid DSQ readout.} (a)-(b) Amplitude of the first (a) and second (b) order DSQ-resonator couplings $\pmb{\gamma}^{(j)}$ [see Eq.~\eqref{eq:gammas}] against $\varphi$ for different dot detunings. 
 We use $\Delta/h=8$~GHz, $E_{so}/h=0.35$~GHz, and $E_0/h=0.9$~GHz reachable in unstrained Ge~\cite{SM}. Also, $\varphi_r=0.16$, $\mu_1=0.2$~meV, $T/h=5$~GHz, and $U_{11}=U_{22}=4U_{12}=1$~meV. 
(c) Dispersive and direct resonator shifts [see Eq.~\eqref{eq:chi_1}]. We assume the optimal scenario for each coupling with $\pmb{\gamma}^{(1)}\perp \textbf{e}_z$ and $\pmb{\gamma}^{(2)}\parallel \textbf{e}_z$, respectively.  (d) Readout infidelity in logarithmic scale of frequency-modulated and longitudinal readout schemes. Inset: SNR against $t_m$ in double-logarithmic scale.
We use $\chi_1/\hbar= \kappa/2=2\pi\times 0.56$~MHz ($1.1$~MHz),  $A= \kappa\sqrt{ \bar{n}/2}$, with $\bar{n}=2$ ($\bar{n}=10$) and  $|\hbar\omega_r-b_{hy}|/h= 0.2$~GHz for dispersive (direct) readout.
For longitudinal readout, by modulating $T=T_0+\delta T \cos(\omega_r t)$ with $T_0/h=2\delta T/h=5$~GHz at $\mu_2-\mu_1=50$~$\mu$eV, $\varphi=0$, and $\pmb{\gamma}^{(1)}\parallel\textbf{e}_z$, we estimate $\delta \eta\approx\delta T \partial_T \eta\approx 0.06$ and $g_z/h\approx 1.26$~MHz.
\label{fig:2}
}
\end{figure}

\paragraph{Hybridized DSQ.--} 
The hopping Hamiltonian $H_\text{1P}$ in Eq.~\eqref{eq:H_hopping_1P} enables different readout schemes. A possible protocol involves physically transferring the spin from the DSQ to the ASQ, e.g. by activating $T$ while decreasing $\epsilon$, and then reading out the ASQ spin via its large direct inductive coupling $\textbf{b}_2^{(2)}$ to the resonator~\cite{lu2025andreev}. 

While experimental advances in shuttling DSQs make this approach promising~\cite{doi:10.1126/science.ado5915,vanRiggelen-Doelman2024}, here we consider the case where the spin remains confined in the  DSQ and  hybridizes to the ASQ by $T_\varphi\lesssim \epsilon_\varphi$, in analogy to flopping-mode qubits~\cite{PhysRevResearch.2.012006,PhysRevResearch.3.013194}.
The ASQ modifies the DSQ's Zeeman energy as 
\begin{equation}
\label{eq:Hybrid-Zeeman}
H_{hy}= \frac{(1-\eta) b_1  \sigma_z}{2}  +\eta \frac{ b_2  {( R_{so}\textbf{e}_z)\cdot\pmb{\sigma}  } }{2} \ , 
\end{equation}
where $S_{so}\sigma_z S_{so}^\dagger= (R_{so} \textbf{e}_z)\cdot\pmb{\sigma}$ includes tilting of $g$-tensors and spin-flip tunneling. 

The DSQ-ASQ hybridization is captured by  $\eta=1/2-\epsilon_\varphi/2E_g\approx T_\varphi^2/\epsilon_\varphi^2$, where $E_g=\sqrt{4T_\varphi^2+\epsilon_\varphi^2}$ is the orbital energy, and mediates the DSQ-resonator coupling   $H_{hy,I}= (\varphi_r X \pmb{\gamma}^{(1)}  + \varphi_r^2 X^2\pmb{\gamma}^{(2)}) \cdot\pmb{\sigma}$, where
\begin{subequations}
\label{eq:gammas}
\begin{align}
\pmb{\gamma}^{(1)} & \approx \eta \frac{{R_{so}\textbf{b}_2^{(1)}}}{2} -{ \eta^{(1)}}\frac{b_1-b_2 R_{so}}{2} \textbf{e}_z    \ , \\
\pmb{\gamma}^{(2)} & \approx\eta \frac{{R_{so}\textbf{b}_2^{(2)}}}{2} -{ \eta^{(2)}}\frac{b_1-b_2 R_{so}}{2} \textbf{e}_z +\frac{ \eta^{(1)} R_{so}\textbf{b}_2^{(1)}}{2}  \ .
\end{align}
\end{subequations}

 The vectors $\pmb{\gamma}^{(j)}$ comprise two distinct contributions: the magnetic susceptibility of the ASQ $\textbf{b}_2^{(j)}= \partial^j_{\varphi} \textbf{b}_2/j$ and the orbital susceptibility $\eta^{(j)}= \partial^j_\varphi\eta/j$. These have a complementary $\varphi$ dependence, opposite for $\pmb{\gamma}^{(1)}$ and $\pmb{\gamma}^{(2)}$, as shown in Fig.~\ref{fig:2}(a)-(b), enabling the selective utilization of different couplings by flux biasing.
 
\paragraph{Readout fidelity.--}
The DSQ-resonator interaction enables an efficient discrimination of spin up and down states. 
The readout is performed by driving the resonator via a weakly-coupled input field $a_\text{in}=Ae^{-i\omega t}/\sqrt{\kappa}$ with amplitude $A$ and coupling rate $\kappa$, yielding  the driving Hamiltonian  $H_d^R= \sqrt{2}A X$ in the frame rotating at the driving frequency $\omega=\omega_r$.
In this frame, to second order in $\varphi_r$, the DSQ-resonator interaction Hamiltonian produces the spin-dependent ac Stark shift
\begin{equation}
\label{eq:chi_1}
H_{hy, I}^{R}= \chi_{1} a^\dagger a \sigma_z \ , \ \text{with} \ \chi_1\approx \frac{\varphi_r^2\left|\pmb{\gamma}^{(1)}\times \textbf{e}_z\right|^2}{2(b_{hy}-\hbar\omega_r)}+ \varphi_r^2\pmb{\gamma}^{(2)}\cdot\textbf{e}_z \ .
\end{equation}
This includes a dispersive~\cite{RevModPhys.93.025005} and direct inductive coupling~\cite{lu2025andreev}, dependent on $\pmb{\gamma}^{(1)}\times \textbf{e}_z$ and $\pmb{\gamma}^{(2)}\cdot \textbf{e}_z$, respectively. We neglect the Lamb shift modifying the DSQ frequency, and introduce the Zeeman field amplitude $b_{hy}=|[(1-\eta) b_1+ \eta b_2 R_{so}]\cdot\textbf{e}_z|$, see Eq.~\eqref{eq:Hybrid-Zeeman}.

Importantly, these contributions are complementary in two ways. Comparing to Fig.~\ref{fig:2}(c), their flux dependence is opposite: at $\varphi=0$, the dispersive and direct couplings are maximal and minimal, respectively, and at $\varphi=\pi/2$, the situation is inverted. We expect $\chi_1/h\sim 1$~MHz, comparable to superconducting qubits~\cite{RevModPhys.93.025005}, enabling fast readout.
Moreover, they show complementary trends depending on the direction of $\pmb{\gamma}$ relative to the Zeeman field: dispersive and direct couplings depend on the transversal and longitudinal components of $\pmb{\gamma}^{(j)}$, respectively. Because the angle $\pmb{\gamma}^{(j)}\cdot \textbf{e}_z$ can be engineered e.g. by aligning  $\textbf{B}$ to the SOI~\cite{PhysRevLett.129.066801,PhysRevB.108.245406} or by electrically tuning $g$-tensors~\cite{Saez-Mollejo2025,john2024two}, we compare optimal scenarios with $\pmb{\gamma}^{(1)}\perp \textbf{e}_z$ and $\pmb{\gamma}^{(2)}\parallel \textbf{e}_z$.

Similar to superconducting qubits, $\chi_1$ imprints a spin dependence to the output field $a_\text{out}$ integrated for a time $t_m$, which enables readout of the DSQ with signal-to-noise-ratio~\cite{PhysRevLett.115.203601,RevModPhys.93.025005}
\begin{equation}
\label{eq:red}
 \text{SNR}= |A| \sqrt{\frac{8t_m}{\kappa}}\left[1-\frac{2}{\kappa t_m}\left(1-e^{-\frac{\kappa t_m}{2}}\cos\left(\frac{\kappa t_m}{2}\right)\right)\right]\ ,
\end{equation}  
where we use the maximal steady-state SNR condition $\chi_1/\hbar=\kappa/2$~\cite{PhysRevA.77.012112}. 
The SNR is approximately related to the readout infidelity by $1-\mathcal{F}=\text{erfc}(\text{SNR}/2)$~\cite{RevModPhys.82.1155,PhysRevApplied.7.054020}. We focus on ideal readouts, and neglect reductions in fidelity coming from limited readout chain efficiency~\cite{hu2025mixed} and improvements by pulse shaping~\cite{PhysRevLett.112.190504,PhysRevApplied.5.011001,PhysRevApplied.6.034008}.

In Fig.~\ref{fig:2}(d), we show that our hybrid DSQ-ASQ system enables high-fidelity readout below microseconds for dispersive and direct couplings (red and blue curves, respectively). In the setup considered, the direct coupling (blue curve) enables faster readout, see Fig.~\ref{fig:2}(c). This is not only because of its larger $\chi_1$, but also because the direct coupling is independent of the resonator-qubit detuning $|\hbar\omega_r-b_{hy}|$, and we operate the setup at larger resonator frequencies $\omega_r$. This permits us to relax stringent conditions imposed by the dispersive regime on the maximal driving amplitudes $A\ll A_\text{crit}=\kappa\sqrt{\bar{n}_\text{crit}/2} $, with critical photon number $\bar{n}_\text{crit}=|\hbar\omega_r-b_{hy}|/4\chi_1\approx 25$~\cite{PhysRevA.79.013819,PhysRevApplied.7.054020} at  $0.2$~GHz detuning and $2$~MHz dispersive shift, and suppresses thermal photons and qubit-induced resonator non-linearities~\cite{kurilovich2025high,connolly2025full}.

Finally, the large first-order  longitudinal component of $\pmb{\gamma}^{(1)}\cdot \textbf{e}_z$, complemented by a harmonic modulation of $\eta\to \eta_0+\delta\eta \cos(\omega_r t)$ obtained by driving $T$, $\epsilon$, or $\varphi$, produces the parametric longitudinal coupling $g_z\cos(\omega_r t) (a^\dagger + a)\sigma_z$, with $g_z\approx\varphi_r \delta \eta \textbf{e}_z\cdot R_{so}\textbf{b}_2^{(1)} /2\sqrt{2} $. We estimate that by modulating $T$,  $g_z/h$ can reach the MHz range also when the DSQ and ASQ are detuned. 
This coupling enables an efficient readout~\cite{PhysRevLett.115.203601,Harpt2025} in the short-time limit with $\text{SNR}_z=|g_z| \sqrt{{8t_m}/{\kappa}}\left[1-{2}\left(1-e^{-{\kappa t_m}/{2}}\right)/{\kappa t_m}\right]$, see black lines in Fig.~\ref{fig:2}(d), offering another fast DSQ-readout option.

\paragraph{Two-particle regime.--}

\begin{figure}
\centering
\includegraphics[scale=0.42]{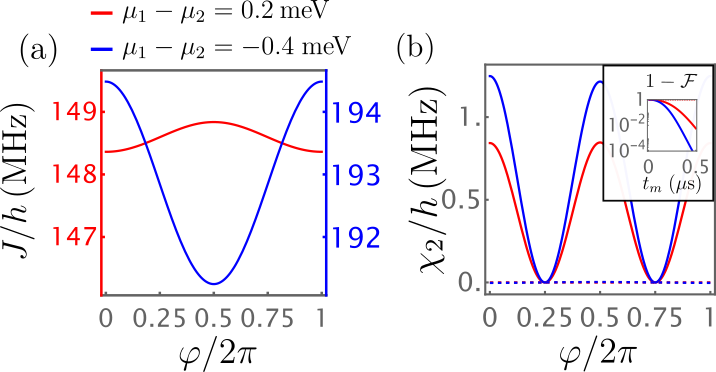}
\caption{ \textbf{ASQ-DSQ exchange interaction.} (a) Phase dependence of $J$ for positive and negative chemical potential shifts. Approaching the (2,0) region (blue curve) or the (0,2) region (red curve) largely increases $J$.
(b) Dispersive (solid lines) and direct (dashed lines) resonator shifts for the same parameters as above. In the inset we show the readout fidelity at $\varphi=0$. We use $T/h=2.5$~GHz and fix $\mu_2=1$~meV. Other parameters as in Fig.~\ref{fig:2}.
\label{fig:3}
}
\end{figure}

We also examine the (1,1) sector of the charge stability diagram where two particles are confined in the double dot, see red dots of Fig.~\ref{fig:1}(b). Filled (hollow) dots represent readout (operational) regions, where the DSQ and ASQ are coupled (decoupled).
Fourth-order perturbation theory of Eqs.~\eqref{eq:HDQD} and~\eqref{eq:HT} results in the exchange-coupled spin Hamiltonian \cite{Saez-Mollejo2025, geyer2022two, bosco2024exchange}
\begin{equation}
\label{eq:anisotropic J}
H_{(1,1)}=\frac{b_1}{2}\sigma_z^{(1)}+\frac{b_2}{2}\sigma_z^{(2)}+ \frac{J}{4}\pmb{\sigma}^{(1)}R_{so}\pmb{\sigma}^{(2)} \ .
\end{equation}
The anisotropic exchange (assuming $U_{11}=U_{22}\equiv U$)
\begin{equation}
J\approx \frac{2T^2}{U_+}+\frac{2T^2}{U_-} -
\frac{2T^2\Delta_\varphi^2}{U_+^2(U_{12}-\Sigma)} +\frac{2T^2\Delta_\varphi^2}{U_-^2(2{U}+3{U}_{12}-\Sigma)}
\end{equation}
comprises a DSQ contribution $\propto T^2/U$ and an ASQ-mediated contribution $\propto T^2\Delta_\varphi^2/U^3$. We introduced $U_{\pm}={U}-U_{12}\pm \epsilon$, $\Sigma=\epsilon_1+\epsilon_2$, and discarded corrections $\propto T^4$ and $\propto b_i b_j$. We expect $J/h\in [0-1]$~GHz in the range reached for DSQs. Because of the phase-dependence of $\epsilon_2$ and $\Delta_\varphi$, $J$  inherits an even $\varphi$-dependence, see Fig.~\ref{fig:3}(a), whose sign and amplitude depend on $\epsilon$.

The system is coupled to the resonator by 
\begin{equation}
H_{(1,1),I}\approx \sum_j \frac{ \varphi_r^j X^j \textbf{b}_2^{(j)}\cdot \pmb{ \sigma}^{(2)}}{2}+\frac{\varphi_r^j X^j J^{(j)}  }{4}\pmb{\sigma}^{(1)}R_{so}\pmb{\sigma}^{(2)} \ ,
\end{equation}
where we identify two distinct contributions: the magnetic modulation of $\textbf{b}_2$ and the charge modulation of $T$ and $\epsilon_2$, which yields a $\varphi$-dependent $J$, with susceptibility $J^{(j)}=\partial_\varphi^jJ/j$. 
We focus on exchange-mediated processes and neglect the DSQ-ASQ spin hybridization  $\propto b_i T^2/U^2$, similar to Eq.~\eqref{eq:Hybrid-Zeeman}.

\paragraph{Exchange-mediated readout.--}
The tuneable anisotropic exchange in Eq.~\eqref{eq:anisotropic J} enables different efficient readout schemes. 
For $b_1\sim b_2$ and $R_{so}\sim 1$, a convenient readout protocol involves first a SWAP gate between DSQ and ASQ, implemented by activating the approximately isotropic exchange $J$ for a time $t_\text{SWAP}=\hbar\pi/J$~\cite{Kandel2019}, followed by fast ASQ readout. 

However, the anisotropic exchange interactions of Ge~\cite{Saez-Mollejo2025} enable exchange-mediated readout protocols. 
By restricting to $J\ll |b_1-b_2|$, $H_{(1,1)}$ reduces to~\cite{nguyen2025single}
\begin{equation}
H_{(1,1)}^\parallel= \frac{b_1}{2}\sigma_z^{(1)}+\frac{b_2}{2}\sigma_z^{(2)}+ \frac{J_\parallel}{4}\sigma_z^{(1)}\sigma_z^{(2)} \ ,
\end{equation}
with a tuneable Ising interaction $J_\parallel=J \textbf{e}_zR_{so}\textbf{e}_z$. 
Neglected flip-flop interactions can induce SWAP dynamics, potentially leading to readout errors. 
These errors are minimized by exchange anisotropies~\cite{PhysRevB.109.085303,geyer2022two,nguyen2025single} and large Zeeman energy differences, which in Ge reach $(b_1- b_2)/h\sim (b_1+b_2)/h\sim 1$~GHz at $200$~mT~\cite{PhysRevLett.128.126803}. Under  these conditions, the SWAP dynamics occur on timescales $\sim h|b_1-b_2|/J^2\gtrsim 10$~$\mu$s at $J/h\lesssim 300$~MHz, longer than the target sub-microsecond  integration time.

Because of $J_\parallel$, the ASQ Stark shift inherits a DSQ-state dependence, yielding the three-body interaction in the rotating frame  $H_{(1,1),I}^R=\chi_2 \sigma_z^{(1)}\sigma_z^{(2)}a^\dagger a $, with
\begin{equation}
\chi_2\approx  \frac{\varphi_r^2 J_\parallel|\textbf{b}_2^{(1)}\times \textbf{e}_z|^2}{16(b_2-\hbar\omega_r)^2 -4J_\parallel^2}+ \frac{\varphi_r^2 J^{(2)}_\parallel}{4} \ .
\end{equation}

With the same parameters as for the hybrid readout, halving $T$ to suppress flip-flop terms, we estimate a dispersive coupling in the MHz range, see Fig.~\ref{fig:3}(b), exceeding the direct coupling and comparable to $\chi_1$.
The dispersive shift $\chi_2$ enables sub-microsecond DSQ readout with reasonable fidelity by keeping the ASQ in its groundstate and detecting the DSQ-state-dependent shift of the resonator frequency. This provides an alternative route to fast and high-fidelity DSQ readout.

\paragraph{Conclusion.--}
We present tunable hybrid couplers composed of DSQs tunnel-coupled to ASQs, which are themselves inductively coupled to resonators. Focusing on germanium-based platforms, we investigate multiple readout schemes in configurations where the coupler is occupied by either one or two particles. In both regimes, we show that our architecture supports fast, high-fidelity DSQ readout within sub-microsecond timescales, enabled by dispersive, direct, or longitudinal resonator coupling. Beyond readout, our ASQ–DSQ coupler can offer a promising route to entangle distant DSQs, offering more flexible and scalable layouts for spin-based quantum processors.

\paragraph{Acknowledgments.--}
We acknowledge support by the EU through the H2024 QLSI2 project, the Army Research Office under Award Number: W911NF-23-1-0110, and NCCR Spin Grant No. 51NF40-180604.  KL acknowledges support by the Laboratory for Physical Sciences through the Condensed Matter Theory Center. AC acknowledges support from the IGNITE project under grant agreement no. 101069515 of the Horizon Europe Framework Programme, as well as the US Army Research Office under Award Number: W911NF-24-2-0043. The views and conclusions contained in this document are those of the authors and should not be interpreted as representing the official policies, either expressed or implied, of the Army Research Office or the U.S. Government. The U.S. Government is authorized to reproduce and distribute reprints for Government purposes notwithstanding any copyright notation herein.

\bibliography{literature}

\clearpage
\newpage
\mbox{~}

\onecolumngrid

\setcounter{equation}{0}
\setcounter{figure}{0}
\setcounter{table}{0}
\setcounter{section}{0}

\renewcommand{\theequation}{S\arabic{equation}}
\renewcommand{\thefigure}{S\arabic{figure}}
\renewcommand{\thesection}{S\arabic{section}}
\renewcommand{\bibnumfmt}[1]{[S#1]}

\begin{center}
  \textbf{\large Supplemental Material: Fast readout of quantum dot spin qubits via Andreev spins}\\[.2cm]
 Michèle Jakob$^{1,*}$, Katharina Laubscher$^{2,*}$, Patrick Del Vecchio$^{1}$,\\ Anasua Chatterjee$^{1}$, Valla Fatemi$^3$, Stefano Bosco$^{1,\dagger}$ \\[.1cm]
   {\itshape $^1$QuTech and Kavli Institute of Nanoscience, Delft University of Technology, Delft, The Netherlands} \\  {\itshape $^2$Condensed Matter Theory Center and Joint Quantum Institute, Department of Physics, University of Maryland, College Park, Maryland 20742, USA}\\
{\itshape $^3$School of Applied and Engineering Physics, Cornell University, Ithaca, NY, 14853, USA  }\\
\end{center}
\maketitle	

\section*{Abstract}
In the Supplemental Material we provide more details on the derivation of the Hamiltonian coupling the Andreev Spin Qubit (ASQ) with the Dot Spin Qubit (DSQ), including the functional dependence of the ASQ parameters on the superconducting phase difference. We also analyze numerically the range of validity of our effective Hamiltonian. Finally, focusing on unstrained germanium heterostructures, where spin-orbit interaction (SOI) is large, we estimate the range of possible parameters that can be reached in state-of-the-art devices.

\section{Effective double-dot Hamiltonian}
\label{sec:DQD}
		
To describe the DSQ--ASQ hybrid system, we start from the Hamiltonian $H=H_{DQD}+H_{L}+H_{L-D2}$ with
\begin{subequations}
\label{eq:HDQD}
\begin{align}
\label{S1}
H_{DQD}&=\sum_{i} \mathbf{c}_{i}^\dagger \left(-\mu_i\sigma_0 +\frac{\mu_B g_i \mathbf{B}\cdot\boldsymbol{\sigma}}{2}\right) \mathbf{c}_{i} -{T}
\left(\textbf{c}_{1}^\dagger \tilde{S}_{so} \textbf{c}_{2} +\textbf{c}_{2}^\dagger \tilde{S}_{so}^\dagger \textbf{c}_{1}\right)+ \sum_{ij,\sigma\sigma'} \frac{U_{ij}}{2}(1-\delta_{ij}\delta_{\sigma\sigma'})n_{i \sigma} n_{j \sigma'}, \\
\label{S2}
H_{L}&=\sum_{\alpha,k} \xi_{\alpha k} \mathbf{d}_{\alpha k }^\dagger  \mathbf{d}_{\alpha k }+\bigg(\sum_{\alpha,k}\Delta_{\alpha } e^{(-1)^\alpha i\varphi/2}d_{\alpha k \uparrow}^\dagger d_{\alpha -k \downarrow}^\dagger +\sum_{k}t_{sc}  \mathbf{d}_{1 k }^\dagger  S_{so}^{sc}\mathbf{d}_{2 k} + \mathrm{H.c.}\bigg), \\
\label{S3}
H_{L-D2}&=\sum_{\alpha,k} t_{\alpha} \mathbf{d}_{\alpha k}^\dagger S_{so}^{\alpha}\mathbf{c}_{2}+  \mathrm{H.c.}
\end{align}
\end{subequations}
Equation~\eqref{S1} describes the bare double quantum dot, where $\textbf{c}_i=(c_{i\uparrow},c_{i\downarrow})$ is the vector of annihilation operators for electrons in dot $i\in\{1,2\}$, $\mu_i$ ($g_i$) is the chemical potential ($g$ tensor) for dot $i$, $\mathbf{B}$ is the magnetic field, $\boldsymbol{\sigma}=(\sigma_x,\sigma_y,\sigma_z)$ is the vector of Pauli matrices acting in spin space, $T$ is the interdot tunneling amplitude, $\tilde{S}_{so}= e^{i\tilde{\theta}_{so} \tilde{\textbf{n}}_{so}\cdot \pmb{\sigma}/2}$ describes a unitary evolution in spin space due to SOI, $n_{i\sigma}=c_{i\sigma}^\dagger c_{i\sigma}$ is the number operator, and $U_{ii}$ ($U_{12}=U_{21}$) are Hubbard intra-dot (inter-dot) interactions. Note that here we directly subtract the diagonal term $\propto (n_{i\sigma})^2=n_{i\sigma}$ from the Hubbard interaction, whereas this contribution is included in the single-particle energy in the main text for convenience.

Equation~\eqref{S2} describes the two superconducting leads, where $\textbf{d}_{\alpha k}=(d_{\alpha k\uparrow},d_{\alpha k\downarrow})$ is the vector of annihilation operators for electrons of momentum $k$ in lead $\alpha\in\{1,2\}$, $\xi_{\alpha k}$ is the onsite energy in lead $\alpha$, $t_{sc}$ is a direct tunneling between the two leads originating from tracing out non-resonant ASQ levels, where we again account for a unitary evolution in spin space $S_{so}^{sc}= e^{i{\theta}_{so}^{sc}{\textbf{m}}_{so}^{sc}\cdot \pmb{\sigma}/2}$ due to SOI, and $\Delta_{\alpha}$ is the effective pairing amplitude in lead $\alpha$. Furthermore, $\varphi$ is the phase difference across the junction. 
In the main text, we also add a resonator with energy $\hbar\omega_r a^\dagger a$ coupled inductively to the ASQ, which modifies the phase difference as $\varphi\to\varphi+\varphi_r(a+a^\dagger)/\sqrt{2}$, where $a^\dagger$, $a$ are creation, annihilation operators for the resonator mode, respectively.

Equation \eqref{S3} describes the coupling between the superconducting leads and the second quantum dot, which together form the ASQ. Here $t_{\alpha}$ is the tunneling amplitude between the dot and lead $\alpha$, where $S_{so}^\alpha= e^{i{\theta}_{so}^\alpha {\textbf{m}}_{so}^\alpha\cdot \pmb{\sigma}/2}$ is once again a unitary evolution in spin space due to SOI.
	
For simplicity, we assume the leads to be symmetric and set $\xi_{1k}=\xi_{2k}\equiv\xi_k$ with $\xi_k=\hbar^2k^2/2m-\mu_{sc}$, $\Delta_{1}=\Delta_{2}\equiv\Delta_0$, and $t_1=t_2\equiv t$. We further consider $\theta_{so}^1=-\theta_{so}^2=\theta_{so}^{sc}/2$ consistent with the setup shown in Fig.~1 of the main text. For simplicity, we set $\textbf{m}_{so}^{1}=\textbf{m}_{so}^{2}=\textbf{m}_{so}^{sc}\equiv\textbf{m}_{so}$ here, but our results can be generalized to arbitrary rotations.
In analogy to the main text, we denote with $\tilde{\textbf{n}}_{so}$ the direction of the SOI between the DSQ and the ASQ, and with $\textbf{m}_{so}$ the direction SOI between the superconducting leads and the ASQ, which is not generally aligned to $\tilde{\textbf{n}}_{so}$. Moreover, in the main text we distinguish between the vector of SOIs $\tilde{\textbf{n}}_{so}$ and the vector ${\textbf{n}}_{so}$, which includes the tilts of quantisation axes of between ASQ and DSQ, i.e. $U_1^\dagger e^{i \tilde{\theta}_{so}\tilde{\textbf{n}}_{so}\cdot \pmb{\sigma}/2} U_2\equiv e^{i {\theta}_{so} \textbf{n}_{so}\cdot \pmb{\sigma}/2}$, where the unitary operators $U_{i}= e^{-i \Theta_i \textbf{n}_i\cdot \pmb{\sigma}/2} $, satisfy $U_i^\dagger (\textbf{b}_i\cdot\pmb{\sigma}) U_i=b_i\sigma_z   $  and align both Zeeman fields $\textbf{b}_{i=1,2}$ to the $z$-direction.\\

We start by focusing on the large-gap limit $\Delta_0\gg U_{ij}$. This allows us to integrate out the superconducting leads and to derive a simple effective Hamiltonian for the ASQ-DSQ hybrid double dot. For this, we first diagonalize the leads via a standard Bogoliubov transformation $d_{\alpha k\uparrow}^\dagger=e^{(-1)^\alpha i \varphi/4}(\cos\phi_k\,\tilde{d}_{\alpha k\uparrow}^\dagger+\sin\phi_k\,\tilde{d}_{\alpha k\downarrow})$, $d_{\alpha -k\downarrow}^\dagger=e^{(-1)^\alpha i \varphi/4}(\cos\phi_k\,\tilde{d}_{\alpha k\downarrow}^\dagger-\sin\phi_k\,\tilde{d}_{\alpha k\uparrow})$ with $\tan 2\phi_k=\Delta_0/\xi_k$ and energy $E_k=\sqrt{\xi_k^2+\Delta_0^2}$. Then, we perform a Schrieffer-Wolff transformation to third order in the tunnel couplings and Zeeman field to obtain an effective low-energy Hamiltonian describing the DSQ-ASQ hybrid, where the leads have now effectively been removed from the problem. This results in the effective double-dot Hamiltonian $H=H_0+H_T$, where 
\begin{subequations}
\label{S5}
\begin{align}
H_0&=\sum_i\mathbf{c}_{i}^\dagger (-\epsilon_i \sigma_0 +\mathbf{b}_i\cdot\boldsymbol{\sigma}/2)\mathbf{c}_{i}- \Delta_\varphi(c_{2\uparrow}^\dagger c_{2\downarrow}^\dagger +c_{2\downarrow} c_{2\uparrow}) +\sum_{ij,\sigma\sigma'} \frac{U_{ij}}{2}(1-\delta_{ij}\delta_{\sigma\sigma'})n_{i \sigma} n_{j \sigma'},\\
H_T&=-{T}
\left(\textbf{c}_{1}^\dagger \tilde{S}_{so} \textbf{c}_{2} +\textbf{c}_{2}^\dagger \tilde{S}_{so}^\dagger \textbf{c}_{1}\right)\ .
\end{align}
\end{subequations}
These are Eqs.~(1) and (2) in the main text. Here, the effective parameters describing the DSQ in dot 1 remain unchanged compared to Eq.~\eqref{S1}, i.e. we have $\epsilon_1=\mu_1$ and $\textbf{b}_1=\mu_B \mathbf{B}g_1$, while the parameters describing the ASQ in dot 2 are modified by the coupling to the superconducting leads as 
\begin{align}
\epsilon_2&=\mu_2 + E_0 \cos(\varphi),\\
\textbf{b}_2&=\mu_B \textbf{B}g_2 - E_{so}\sin(\varphi)\textbf{m}_{so},  \\
\Delta_\varphi &= \Delta\cos(\varphi/2) ,
\end{align}
where
\begin{align}
 E_0&=2t^2 t_{sc}  \cos(\theta_{so}^{sc}) \int dk \frac{\sin^2(\phi_k)}{E_k^2}\approx \Gamma \frac{ t_{sc} }{\Delta_0} \cos(\theta_{so}^{sc})  \ , \\
  E_{so}&= 4t^2 t_{sc}  \sin(\theta_{so}^{sc}) \int dk \frac{\sin^2(\phi_k)}{E_k^2}\approx 2 \Gamma \frac{ t_{sc} }{\Delta_0} \sin(\theta_{so}^{sc})  \ , \\
  \Delta&=2t^2 \int dk \frac{\sin(\phi_k)}{E_k}\approx 2 \Gamma   \ .
\end{align}
In the approximation, we have assumed a constant density of states $\rho$ for the leads and introduced the tunneling rate $\Gamma= \pi \rho t^2$. We also neglect the small additional corrections arising form third-order perturbation theory:
\begin{equation}
  2 t^2 \int \frac{d k}{E_k^2} \approx \frac{2\Gamma}{\Delta_0}
\end{equation}
that renormalize $\epsilon_2\to \epsilon_2(1-2\Gamma/\Delta_0)$ and $\textbf{b}_2\to \textbf{b}_2(1-2\Gamma/\Delta_0)$ because of hybridization to the leads. These corrections are consistent with the generalized atomic limit substitutions $1/(1+2\Gamma/\Delta_0)$ in the small tunnel limit~\cite{PhysRevB.107.115407}. Similarly, in this limit we neglect the same renormalization of the Hubbard energy $U_{22}$.  
	
While working in the large-gap limit has allowed us to derive a very simple effective Hamiltonian for the DSQ-ASQ hybrid, experimentally realistic systems are often in the opposite limit $U_{ij}\gg\Delta_0$. Therefore, in Sec.~\ref{sec:ED}, we will additionally study a simplified version of $H$ in the opposite regime $U_{ij}\gg\Delta_0$ using exact numerical diagonalization.

We also note that the phase dependence of the ASQ parameters, as well as the linear dependence of $E_{so}$ on $\Gamma$, $t_{sc}$, and $\theta_{so}^{sc}$ when these quantities are small, was also confirmed by NRG calculations~\cite{PhysRevLett.131.097001}. A more sophisticated numerical analysis is required to find the precise dependence on $\Delta_0$, especially in the regime $\Gamma\gtrsim \Delta_0$, where our Schrieffer-Wolff perturbation theory fails.

\section{Estimation of ASQ parameters in Germanium}
Here, we estimate a possible range of the relevant parameters for unstrained germanium (Ge) channels confined along the $z$ direction.
The 2D Hamiltonian for an unstrained Ge hole gas and long wavelengths up to third order in momentum $k_{x,y}$ is~\cite{PhysRevB.103.125201}
\begin{equation}\label{eq:Heff}
   H= \frac{\hbar^2 (k_x^2+k_y^2)}{2m^*} +i\sigma_+(\beta_1 k_- -\beta_2 k_+^3 + \beta_3 k_-k_+k_-) + \text{H.c.} \ ,
\end{equation}
where $k_\pm=k_x\pm ik_y$, $\sigma_\pm=(\sigma_x\pm\sigma_y)/2$, $m^*$ is the effective 2D mass of the holes, and $\beta_{1,2,3}$ characterize the SOI.
The unstrained nature of the channel, demonstrated in current experiments~\cite{Desamaterial}, enables a large enhancement of the SOI, which is required for ASQs in the absence of magnetic fields.
Because $\beta_3/\beta_2\approx(\gamma_3-\gamma_2)/(\gamma_3+\gamma_2)\approx 0.14$, we neglect anisotropies in the effective Hamiltonian here and set $\beta_3=0$. We introduce here the material-dependent Luttinger-Kohn parameters $\gamma_{1,2,3}$.   This term results in an enhanced SOI for confinement along the [110]-direction.
We also note that typically $\beta_1\sim 0$ in 2D heavy hole (HH) gasses and neglect it here. Larger SOI is expected in light hole (LH) gasses~\cite{del2025fully}.

Here, we focus on the buried unstrained Ge channel demonstrated experimentally in Ref. \cite{Desamaterial}, which is characterized by $\beta_2\approx 850\,\text{meV}\,\text{nm}^3$ and $m_0/m^*\approx 13$, where $m_0$ is the bare electron mass. We note that current strained Ge quantum wells have a dramatically lower SOI, roughly 1-2 orders of magnitude smaller ($\beta_2\sim 0.9\,\text{meV}\,\text{nm}^3$ at a $1\,\text{V}/\mu\text{m}$ gate field), thus leading to a much smaller spin splitting.
This estimation for $\beta_2$ comes from a third-order Schrieffer-Wolff transformation starting from the 6-dimensional Luttinger-Kohn Hamiltonian including HH, LH, and split-off holes (SO), see the Supplemental Material of Ref.~\cite{Desamaterial} for more details on the procedure, which leads to the Hamiltonian in Eq.~\eqref{eq:Heff}. The amplitude of the SOI is explicitly given by
\begin{equation}
\beta_2=\left(\frac{\hbar^2}{2m_0}\right)^2\sum_j{\frac{T_{1,j}^\text{x}\mu_{1,j}^* + \mu_{1,j}T_{1,j}^{\text{x}*}}{\left(E_1^\text{H}-E_j^\eta\right)}} - \frac{1}{2}\left(\frac{\hbar^2}{2m_0}\right)^3\sum_{j,j^\prime}{\frac{T_{1,j}^\text{x}T_{j,j^\prime}^\eta T_{1,j^\prime}^{\text{x}*}}{\left(E_1^\text{H} - E_j^\eta\right)\left(E_1^\text{H} - E_{j^\prime}^\eta\right)}},
\end{equation}
\noindent where $E_{l=1}^\text{H}$ and $E_j^\eta$ are the (spin-independent) energies at $k_x=k_y=0$ of the HH ground state and of the $j$th LH/SO subband respectively ($\eta$ labels subbands of pseudo-spin $1/2$, which can be either LH-like or SO-like). The coefficients $T$ and $\mu$ are given by
\begin{align}
    \mu_{l,j} &= \frac{3}{2}\braket{f^h_l|\gamma_2 + \gamma_3|f_j^\circ}, \\
    T_{l,j}^\text{x} &= -\frac{3i}{\sqrt{2}}\bra{f_l^h}\left(u_+\ket{f^z_j} + \frac{7\sqrt{6}}{12}[q,k_z]\ket{f^\ell_j}\right), \\
    T_{j,j^\prime}^\eta &= -\frac{3i}{\sqrt{2}}\left(\braket{f^\circ_j|u_+|f^z_{j^\prime}} - \braket{f^z_j|u_-|f^\circ_{j^\prime}}\right) - 5i\braket{f^\ell_j|[q,k_z]|f^\ell_{j^\prime}}, \\
    u_\pm &= \{\gamma_3,k_z\} \pm [\kappa,k_z],
\end{align}
\noindent where $\braket{z|f^h}\equiv f^h(z)$ is the envelope function of HH subbands, and $\ket{f^\ell_j}$ is the envelope function of the LH spinor component of the $j$th LH/SO subband. $\{A,B\} = AB + BA$ is the anti-commutator. We also introduce the Luttinger-Kohn parameters $\kappa$ and $q$. Finally, the envelopes $\ket{f^\circ}$ and $\ket{f^z}$ are linear combinations of $\ket{f^\ell}$ and $\ket{f^s}$:
\begin{subequations}
\begin{align}
    \ket{f_j^z} &\equiv \frac{1}{\sqrt{3}}\left(\sqrt{2}\ket{f_j^\ell} - \ket{f_j^s}\right), \\
    \ket{f_j^\circ} &\equiv \frac{1}{\sqrt{3}}\left(\ket{f_j^\ell} + \sqrt{2}\ket{f_j^s}\right),
\end{align}
\end{subequations}
\noindent where $\ket{f^s_j}$ is the SO spinor component of the $j$th LH/SO subband. The effective mass is calculated similarly to $\beta_2$, but its exact description in terms of the envelopes and subband energies at $k_x=k_y=0$ only requires a second-order Schrieffer-Wolff transformation. The inverse effective mass is given by
\begin{equation}
   \frac{m_0}{m^*} = \braket{f^h_1|\gamma+\gamma_2|f^h_1} - \frac{\hbar^2}{2m_0}\left(\sum_l{\frac{T_{1,l}^\text{H}T_{l,1}^\text{H}}{E_1^\text{H} - E_l^\text{H}}} + \sum_j{\frac{T_{1,j}^\text{x}T_{1,j}^{\text{x}*}}{E_1^\text{H} - E_j^\eta}}\right),
\end{equation}
\noindent where
\begin{equation}
    T_{l,l^\prime}^\text{H} = -\frac{3i}{2}\braket{f^h_l|[q,k_z]|f^h_{l^\prime}}.
\end{equation}

The envelope functions $\ket{f}$ and associated energies are computed by diagonalizing the $6$-band $k\cdot p$ Hamiltonian evaluated at $k_x=k_y=0$. Using the basis ordering
\begin{equation}
    \{\text{LH}+,\text{SO}+,\text{HH}+,\text{LH}-,\text{SO}-,\text{HH}-\},
\end{equation}
\noindent where $\pm$ labels the positive and negative spin projection along the $z$-direction, the $k\cdot p$ Hamiltonian is given by
\begin{subequations}
  \begin{align}\label{H0}
    H_0^{k\cdot p} &= \begin{bmatrix}
      H_{\sigma=+} & 0 \\
      0 & H_{\sigma=-}
    \end{bmatrix}, \\
    H_\sigma &= H_\sigma^k + H_\sigma^\varepsilon + V,
  \end{align}
\end{subequations}
\noindent where
\begin{subequations}
  \begin{align}
    H_\sigma^k &= \frac{\hbar^2}{2m_0}\begin{bmatrix}
    -k_z\gamma_+k_z & 2\sqrt{2}\sigma k_z\gamma_2k_z & 0 \\
    & -k_z\gamma_1 k_z & 0 \\
    \dagger & & -k_z\gamma_-k_z
    \end{bmatrix}, \\
    H_\sigma^\varepsilon &= a_v\text{Tr}\,\varepsilon + b\cdot\delta\varepsilon\begin{bmatrix}
    -1 & \sqrt{2}\sigma & 0 \\
    & 0 & 0 \\
    \dagger & & 1
    \end{bmatrix}, \\
    V &= \mathcal{E}_{\Gamma_5^+} + \frac{\delta_0}{3} + eF_zz - \delta_0\begin{bmatrix}
    0 & 0 & 0 \\
    & 1 & 0 \\
    \dagger & & 0
    \end{bmatrix}.
  \end{align}
\end{subequations}

\noindent Here, $\gamma_\pm = \gamma_1\pm 2\gamma_2$ are combinations of Luttinger parameters, $a_v$ and $b$ are strain deformation potentials, $\delta\varepsilon = \varepsilon_{xx} - \varepsilon_{zz} \approx 1.67\varepsilon_{xx}$, where $\varepsilon_{ij}$ is the strain tensor, $\delta_0$ is the bulk split-off gap, and $\mathcal{E}_{\Gamma_5^+}$ is the valence band edge energy without SOI, and $F_z$ is an electric field applied perpendicular to the 2D plane. We consider $F_z=0.05\,\text{V}/\mu\text{m}$ and $\varepsilon_{xx}=-0.018\%$ in the Ge layer, which is consistent with the experimental values reported in Ref.~\cite{Desamaterial}. The strain in the SiGe barrier is calculated assuming pseudomorphic growth. The energy band offsets between Ge and SiGe, the deformation potential constant $b$, the $\gamma_i$ Luttinger parameters, and the $g$-factor parameters $\kappa$ and $q$ are parameterized as in Ref.~\cite{Desamaterial}.
We stress that each material parameter ($\gamma_1$, $a_v$, $\mathcal{E}_{\Gamma_5^+}$, etc.) is a function of the position $z$ in the heterostructure. Because of this spatial dependence,  these parameters do not commute with $k_z$ and we treat them as diagonal operators in  Eq.~\eqref{H0}. The $k_z$ operators are implemented using a finite difference scheme with a $\delta z = 0.01\,\text{nm}$ mesh spacing. \\

We consider a range of devices with typical experimental parameters for Ge, i.e. a channel along the $y$ direction of length $l_y\in[120-200]$~nm, corresponding to the orbital energy gap $\omega_y/2\pi=\hbar/2\pi m^*l_y^2\in [16-6]$~GHz, and with lateral confinement energy $\hbar\omega_x=\hbar^2/m^*l_x^2\in [0.4-2.5]$~meV, corresponding to a confinement length $l_x\in[50-20]$~nm.

The SOI angle $\theta_{so}^{sc}\approx l_y/l_{so}^{sc}$ is given by the ratio of the long confinement direction $l_y$ and the SOI length $l_{so}^{sc}=\hbar/m^* v_{so}^{sc}$, which depends on the effective mass $m^*$ and SOI velocity $v_{so}^{sc}$ of the channel. To estimate this in the unstrained Ge channel, we project the cubic Rashba SOI onto the ground state in the lateral confinement $x$ direction, with spatial extension $l_x$, resulting in $\hbar v_{so}^{sc}\approx 3\beta_2/2l_x^2\in [0.5-3.2] $~meV/nm.
This results in $l_{so}^{sc}\in[5.8-0.9]$~$\mu$m, and consequently in $\theta_{so}^{sc}\in[0.02-0.21]$.

We note that the SOI in this case is predominantly aligned in the lateral confinement $x$ direction,  i.e. $\textbf{m}_{so}\approx \textbf{e}_x$, although tilts of the order of $\beta_3/\beta_2$ are expected if the confinement direction is misaligned to the crystal axis. Large tilts can also originate because of inhomogeneous shear strain~\cite{PhysRevLett.131.097002}.

We emphasize that this is a conservative estimate for $v_{so}^{sc}$ and $\theta_{so}^{sc}$ that does not account for the additional increase of linear SOI $\beta_1$ that is caused by lateral confinement~\cite{PhysRevB.104.115425} and inhomogeneous gate-induced strain~\cite{PhysRevLett.131.097002}. These effects are known to drastically  enlarge the resulting SOI.

To estimate the ASQ energies, we consider a Ge ASQ proximitized with platinum germanosilicide (PtGeSi) leads, where $\Delta_0\approx 75$~$\mu$eV ($\Delta_0/h\approx 18$~GHz)~\cite{Lakic2025}.  We restrict the tunneling hybridization to a fraction of the gap  $\Gamma\in [10-20]$~$\mu$eV ($\Gamma/h\in [2.4-4.8]$~GHz), where the results of our perturbative model are applicable. However, we expect our results to remain qualitatively valid for a broader range of tunnelings, including the limit $\Gamma\gtrsim \Delta_0$ which is reachable in current experiments in hybrid Ge systems~\cite{Lakic2025}, see the discussion in Sec.~\ref{sec:ED} and, e.g., NRG calculations~\cite{PhysRevLett.131.097001}. 
Larger superconducting gaps and tunnelings are enabled by utilizing other superconductors such as aluminum or niobium.
The tunneling $t_{sc}$ through non-resonant states is required for ASQs and depends on the contributions of all excited states. We assume it to be of the same order as $\Gamma$, and setting $t_{sc}=\Gamma$, we find the range $E_{so}/h\in[10-600]$~MHz and $E_{0}/h\in[300-1300]$~MHz.

In the main text, we consider ASQ parameters that are well within the range discussed here. In paricular, we use $E_{so}/h=350$~MHz, $E_0/h=900$~MHz, $\Delta/h=8$~GHz, that are obtained for unstrained Ge wells proximitized by PtGeSi with $\Gamma/h=t_{sc}/h=4$~GHz, $l_x=20$~nm, and $l_y=180$~nm.

\section{Effective many-body Hamiltonian} 
	
Starting from Eq.~\eqref{S5}, we can write down the full many-body Hamiltonian for the two dots (ASQ and DSQ) occupied by a maximum of 4 particles. This gives us an effective $16\times 16$ Hamiltonian that splits into two decoupled blocks corresponding to even and odd total parity. The structure of this Hamiltonian resembles that of a conventional double quantum dot with the important difference that the $(n,0)$ and $(n,2)$ (for $n\in\{0,1,2\})$ charge sectors are coupled by the effective pairing $\propto\Delta_\varphi$ on dot 2.

We can write the effective Hamiltonian as
\begin{equation}\label{eq:Hfull_sm}
    H=\left(
    \begin{array}{ccccc}
       H_0  & 0 & H_{02} & 0 & 0  \\
        0 & H_{1} & 0 & H_{13} & 0  \\
        H_{02}^\dagger & 0 & H_2 & 0 & H_{24}   \\
        0 & H_{13}^\dagger & 0 & H_3 &  0 \\
        0 & 0 & H_{24}^\dagger & 0 & H_4  \\
    \end{array}
    \right) \ ,
\end{equation}
where $H_{n}$ in the diagonal represents the Hamiltonian of the $n$th charge sector and $H_{nn'}$ couples charge sectors $n$ and $n'$ by the superconducting correlations.
Explicitly, 
\begin{subequations}
    \begin{align}
        H_0&=0 \ , \\
        H_1&= \left(
\begin{array}{cc}
-\epsilon_1 \sigma_0+{\textbf{b}_1\cdot \pmb{\sigma}}/{2} & -T \tilde{S}_{so} \\
-T \tilde{S}_{so}^\dagger & -\epsilon_2\sigma_0+{\textbf{b}_2\cdot \pmb{\sigma}}/{2}
\end{array}
\right) \ , \\
H_2&=\left( 
\begin{array}{ccc}
 U_{11}-2\epsilon_1    & 0& \Theta \\
  0   &  U_{22}-2\epsilon_2 &  0 \\
   \Theta^\dagger   & 0 & (U_{12}-\epsilon_1-\epsilon_2)\sigma_0^{(1)}\sigma_0^{(2)}+\textbf{b}_1\cdot\pmb{\sigma}^{(1)}\sigma_0^{(2)}/2+\sigma_0^{(1)}\textbf{b}_2\cdot\pmb{\sigma}^{(2)}/2
\end{array}
\right) \ , \\
\Theta&=T \left(
\begin{array}{cccc}
 -i n_+ \sin \left(\tilde\theta _{\text{so}}/2\right) & -\cos \left(\tilde\theta _{\text{so}}/2\right)+i n_z \sin \left(\tilde\theta _{\text{so}}/2\right) & \cos \left(\tilde\theta _{\text{so}}/2\right)+i n_z \sin \left(\tilde\theta _{\text{so}}/2\right) & i n_- \sin \left(\tilde\theta _{\text{so}}/2\right) \\
 i n_+ \sin \left(\tilde\theta _{\text{so}}/2\right) & -\cos \left(\tilde\theta _{\text{so}}/2\right)-i n_z \sin \left(\tilde\theta _{\text{so}}/2\right) & \cos \left(\tilde\theta _{\text{so}}/2\right)-i n_z \sin \left(\tilde\theta _{\text{so}}/2\right) & -i n_- \sin \left(\tilde\theta _{\text{so}}/2\right) \\
\end{array}
\right)\ , \\
H_3&= \left(
\begin{array}{cc}
(U_{22}+2U_{12}-2\epsilon_2-\epsilon_1) \sigma_0+{\textbf{b}_1\cdot \pmb{\sigma}}/{2} & T \tilde{S}_{so}^\dagger \\
T \tilde{S}_{so} & (U_{11}+2U_{12}-2\epsilon_1-\epsilon_2) \sigma_0+{\textbf{b}_2\cdot \pmb{\sigma}}/{2}
\end{array}
\right) \ , \\
H_4&= U_{11}+U_{22}+4U_{12}-2\epsilon_1-2\epsilon_2\ .
    \end{align} 
\end{subequations}
Here, $n_\pm=(\tilde{\textbf{n}}_{so})_1\pm i (\tilde{\textbf{n}}_{so})_2$, $n_z=(\tilde{\textbf{n}}_{so})_3$. If we include superconducting correlations in both dots $\Delta^{(1)}$ and $\Delta^{(2)}$, we find
\begin{subequations}
    \begin{align}
    H_{02}&=\left(\Delta^{(1)},\Delta^{(2)},0,0,0,0 \right) \ , \\
H_{13}&= \text{diag}\left(\Delta^{(2)},\Delta^{(2)},\Delta^{(1)},\Delta^{(1)} \right) \ , \\
H_{24}&=\left(\Delta^{(2)},\Delta^{(1)},0,0,0,0 \right)^T \ .
\end{align}
\end{subequations}
In this work, only dot 2 is proximitized, so we set $\Delta^{(1)}=0$ and $\Delta^{(2)}=\Delta_\varphi$.

Starting from the full Hamiltonian given here, the effective Hamiltonians for the different charge sectors given in Eqs.~(4) and (9) of the main text are then readily obtained by standard Schrieffer-Wolff perturbation theory. 
    
\section{Exact diagonalization in the zero-bandwidth limit}\label{sec:ED}
\begin{figure*}[tb]
	\centering
	\includegraphics[width=0.9\columnwidth]{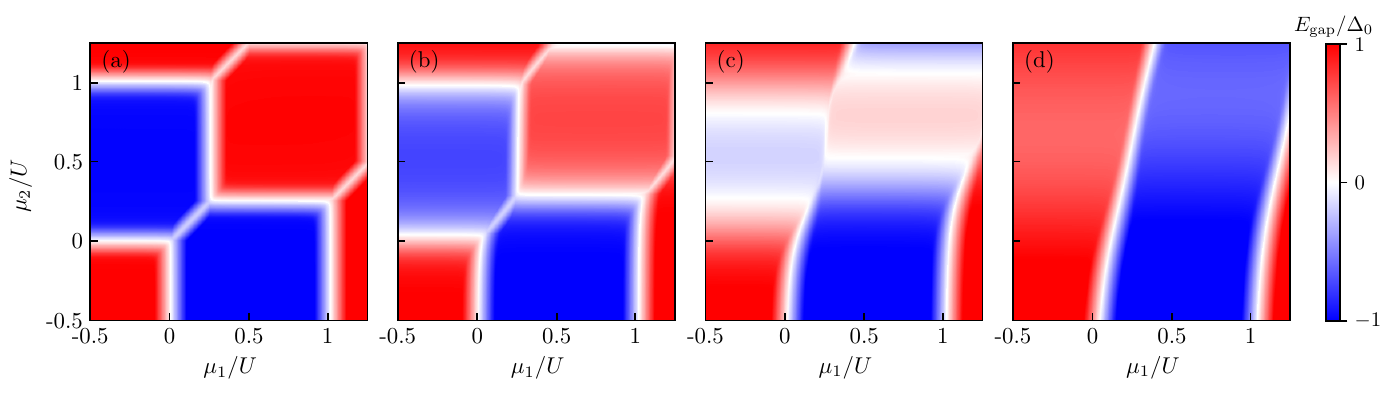}
	\caption{Charge stability diagrams obtained from exact numerical diagonalization of $H_{DQD}+H_L+H_{L-D2}$ using the zero-bandwidth approximation for the superconducting leads. We have set $U_{11}=U_{22}=4U_{12}\equiv U$ and (a) $t=2t_{sc}=0.01U$, (b) $t=2t_{sc}=0.05U$, (c) $t=2t_{sc}=0.1U$, (d) $t=2t_{sc}=0.2U$. The other parameters are $\Delta_0=0.1U$, $\varphi=0$, $\mu_{sc}=0$, $T=0.01U$, $\mathbf{B}=0$, $\tilde\theta_{so}=\theta_{so}^1=-\theta_{so}^2=\theta_{so}^{sc}/2=\pi/5$, $\tilde{\textbf{n}}_{so}=\textbf{m}_{so}^{1}=\textbf{m}_{so}^{2}=\textbf{m}_{so}^{sc}=(0,0,1)$. The color indicates the size of the gap between the ground state manifold and the first excited state. We have multiplied the excitation energies in the doublet sector by a factor of $-1$ such that different colors are assigned to sectors with singlet (red) and doublet (blue) ground states. Note that we have ignored the small splitting between singlet and triplet states in the (1,1) sector arising due to the finite $T$; the excitation gap indicated for this sector corresponds to the gap between the highest-energy triplet state and the next excited state.
    }
	\label{fig:charge_stability_diagram}
\end{figure*}

\begin{figure}[tb]
	\centering
	\includegraphics[width=0.65\columnwidth]{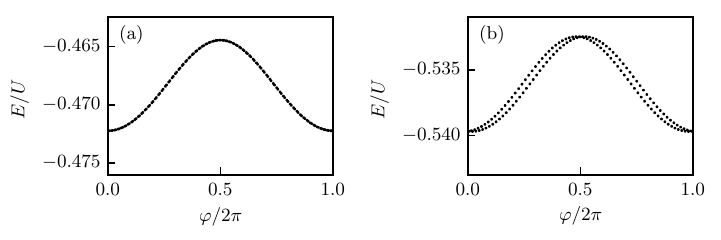}
	\caption{Phase dependence of lowest-energy states in the (1,0) sector obtained from exact numerical diagonalization of $H_{DQD}+H_L+H_{L-D2}$ using the zero-bandwidth approximation for the superconducting leads. We have set $U_{11}=U_{22}=4U_{12}\equiv U$ and (a) $T=0.01U$, (b) $T=0.15U$. The other parameters are $\mu_1=0.25U$, $\mu_2=0$, $\Delta_0=0.1U$, $\mu_{sc}=0$, $t=2t_{sc}=0.05U$, $\mathbf{B}=0$, $\tilde\theta_{so}=\theta_{so}^1=-\theta_{so}^2=\theta_{so}^{sc}/2=\pi/5$, $\tilde{\textbf{n}}_{so}=\textbf{m}_{so}^{1}=\textbf{m}_{so}^{2}=\textbf{m}_{so}^{sc}=(0,0,1)$.
    }
	\label{fig:phase_dependence_10}
\end{figure}

\begin{figure}[tb]
	\centering
	\includegraphics[width=0.65\columnwidth]{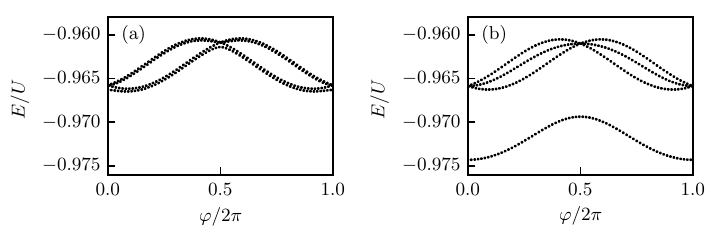}
	\caption{Phase dependence of lowest-energy states in the (1,1) sector obtained from exact numerical diagonalization of $H_{DQD}+H_L+H_{L-D2}$ using the zero-bandwidth approximation for the superconducting leads. We have set $U_{11}=U_{22}=4U_{12}\equiv U$ and (a) $T=0.01U$, (b) $T=0.04U$. The other parameters are $\mu_1=\mu_2=0.5U$, $\Delta_0=0.1U$, $\mu_{sc}=0$, $t=2t_{sc}=0.05U$, $\mathbf{B}=0$, $\tilde\theta_{so}=\theta_{so}^1=-\theta_{so}^2=\theta_{so}^{sc}/2=\pi/5$, $\tilde{\textbf{n}}_{so}=\textbf{m}_{so}^{1}=\textbf{m}_{so}^{2}=\textbf{m}_{so}^{sc}=(0,0,1)$.
    }
	\label{fig:phase_dependence_11}
\end{figure}

To obtain a simple model for which the many-body problem can be solved exactly, we make use of the so-called zero-bandwidth approximation (ZBA) for the superconducting leads~\cite{SeoaneSouto2024}. Within this approximation, the superconducting leads are replaced by just a single site each, leading to a simplified version of the Hamiltonian terms given in Eqs.~\eqref{S2} and \eqref{S3}:
\begin{align}
H_{L}^{ZBA}&=-\mu_{sc}\sum_{\alpha}  \mathbf{d}_{\alpha }^\dagger  \mathbf{d}_{\alpha }+\left(\Delta_0\sum_{\alpha} e^{(-1)^\alpha i\varphi/2}d_{\alpha \uparrow}^\dagger d_{\alpha \downarrow}^\dagger+t_{sc} \mathbf{d}_{1 }^\dagger  S_{so}^{sc}\mathbf{d}_{2}  + \mathrm{H.c.}\right),\\
H_{L-D2}^{ZBA}&=t\sum_{\alpha} \mathbf{d}_{\alpha}^\dagger S_{so}^{\alpha}\mathbf{c}_{2}+  \mathrm{H.c.},
\end{align}
where $\textbf{d}_{\alpha}=(d_{\alpha \uparrow},d_{\alpha \downarrow})$. Here, we have again assumed symmetric leads. The resulting many-body Hamiltonian takes the form of a $256\times 256$ matrix (4 dots occupied by a maximum of 8 particles), which can be diagonalized numerically. This approach allows us to go beyond the large-gap regime $\Delta_0\gg U_{ij}$, where our analytical calculations were performed (see Sec.~\ref{sec:DQD}). In Fig.~\ref{fig:charge_stability_diagram}, we show examples of charge stability diagrams for the DSQ-ASQ hybrid obtained for $\Delta_0=0.1U$ (here $U_{11}=U_{22}=4U_{12}\equiv U$) at zero phase difference $\varphi=0$ and zero magnetic field $\mathbf{B}=0$. In the weak-coupling limit $t\ll U$ , the charge stability diagram takes the form of that of a conventional double quantum dot, see Fig.~\ref{fig:charge_stability_diagram}(a). As the tunnel coupling is increased, the quasiparticles in the superconducting leads begin to screen the electron spin on dot 2 for the (0,1) and (1,1) sectors, see Figs.~\ref{fig:charge_stability_diagram}(b) and (c). Nevertheless, the doublet ground state for dot 2 still survives in some regions of parameter space. Finally, once the tunnel-coupling reaches a critical value, the spin on dot 2 is fully screened, resulting in a singlet state on dot 2 regardless of the dot energy, see Fig.~\ref{fig:charge_stability_diagram}(d).
	
In Fig.~\ref{fig:phase_dependence_10}, we additionally show the phase dependence of the lowest many-body energy levels deep in the (0,1) sector. We observe that an effective phase-dependent spin splitting is induced for the electron confined to the DSQ due to hybridization with the ASQ if the DSQ-ASQ tunneling $T$ becomes large enough, $T\lesssim \epsilon$.

The phase dependence of the lowest energy levels in the (1,1) sector is shown in Fig.~\ref{fig:phase_dependence_11}. Here, we find that a phase-dependent splitting between singlet and triplet states is obtained due the effective exchange between the spin confined in the DSQ and the spin confined in the ASQ. This behavior is qualitatively consistent with the effective exchange Hamiltonian presented in the main text.

We note that while the ZBA has allowed us to work with a very simple numerically tractable model, the applicability of this model is limited due to the absence of a continuum in the superconducting leads. Nevertheless, for a single dot coupled to two superconducting leads, it has been shown that the ZBA qualitatively captures many features of the full model~\cite{PhysRevB.68.035105}, including the singlet-doublet transition. In the case of a double-dot coupled to two leads in parallel or in series, it was found that the limitations of the ZBA become more severe~\cite{PhysRevB.110.134506}. However, in our case, only dot 2 is directly coupled to the SC leads, such that we expect the ZBA to give reasonable results at least as long as the interdot tunneling $T$ is not too large. More sophisticated numerical techniques such as, e.g., NRG or a multi-site extension of the ZBA could be used to verify the results presented here at a more quantitative level.

\end{document}